\documentclass[amsmath,amssymb,aps,pra,twocolumn]{revtex4-2}
\usepackage{tikz}
\usepackage{graphicx}
\usepackage{amssymb,amsfonts,amsmath,amsthm}
\usepackage{braket}
\usepackage{hyperref}
\usepackage{color}
\usepackage{soul}
\usepackage{changes}

\usepackage{enumerate,enumitem}
\usepackage{comment}
\usepackage{hyperref}
\usepackage{eqnarray}
\usepackage{bbold}
\usepackage{optidef}
\usepackage{booktabs}

\def\vect{\vec{t}}
\def\vecsigma{\vec{\sigma}}
\def\vecs{\vec{s}}
\def\vecr{\vec{r}}

\def\Id{\mathbb{1}}

\def\Tr{\mathrm{Tr}}

\def\N{\mathcal{N}}

\def\opt{\mathrm{opt}}
\def\cl{\mathrm{cl}}
\def\avgd{\braket{\bar{d}\,}}
\def\mw{\mathbb{w}}
\def\mR{\mathbb{R}}

\DeclareMathOperator*{\argmin}{argmin}

\usepackage{xcolor}
\begin{document}

\title{Challenges in certifying quantum teleportation: moving beyond conventional fidelity benchmark}

\author{D. G. Bussandri}
\email{diegogaston.bussandri@uva.es (corresponding author)}
\affiliation{Departamento de Física Teórica, Atómica y Óptica, Universidad de Valladolid, 47011 Valladolid, Spain}

\author{G. M. Bosyk}
\email{gbosyk@fisica.unlp.edu.ar}
\affiliation{Instituto de F\'isica La Plata, UNLP, CONICET, La Plata, Argentina}
\author{F. Toscano}%
\email{toscano@if.ufrj.br}
\affiliation{Instituto de Física, Universidade Federal do Rio de Janeiro, Rio de Janeiro, Brazil}

\date{\today}

\begin{abstract}
The conventional certification method for quantum teleportation protocols relies on surpassing the highest achievable classical average fidelity between target and teleported states. Our investigation highlights the limitations of this approach: inconsistent conclusions can be obtained when it is considered different distance measures in the quantum state space, leading to contradictory interpretations. 
In particular, this behavior is manifested when modeling a very common noisy experimental scenario, in which the resource state takes the form of a Werner state generated by the influence of a depolarizing channel acting on the Bell state resource. Two additional noise models, based on amplitude-damping channel, are also analyzed. Our work, therefore, stresses the necessity of new certification methods for quantum teleportation.
\end{abstract}

\maketitle

\section{Introduction}

Quantum teleportation~\cite{Bennett1993,Pirandola2015} is a fundamental protocol in the field of quantum information. 
%
%In the general case, the teleportation protocol consists of a series of local operations applied over a composite global quantum system, assisted by a classical communication channel, which aims to transfer an unknown state $\rho^a$ from one quantum system $a$ to another $B$. 
%
In the general case, it consists of a series of local operations applied over a composite global quantum system, assisted by a classical communication channel, which aims to transfer an unknown state $\rho^a$ from one quantum system $a$ to another $B$. In the ideal case, the sender (Alice) succeeds with probability one in transmitting the state $\rho^a$ of her qubit system $a$ to the target qubit $B$ operated by the receiver (Bob). This protocol relies on preparing a --maximally entangled-- Bell state, usually referred to as the resource state, in the joint system composed of an auxiliary qubit system $A$ and the target system $B$. This step is followed by Alice's measurement: a projection onto the Bell basis $\{\ket{\Phi_i}\}_{i=1}^4$ over qubits $a$ and $A$. Finally, Alice classically communicates its measurement result to Bob who corrects his state $\rho^{B|i}$ by applying local Pauli gates $U_i$ conditioned on Alice's measurement, that is, $\rho^{B}_i= U_i\rho^{B|i}U_i$, resulting in $\rho^B_i = \rho^a$ for all $i\in[1,4]$.

\begin{figure}
    \centering
    \includegraphics[width=0.45\textwidth]{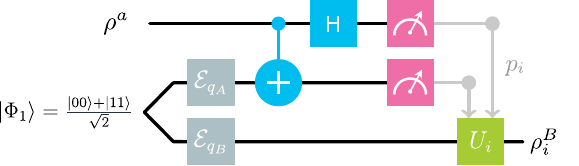}
     \caption{Diagram of quantum teleportation with resource qubits affected by noise via channels $\mathcal{E}{q_A}$ and $\mathcal{E}{q_B}$ on the Bell state $\ket{\Phi_1}$. We consider in this work three different noises: \textit{depolarizing} channel $\mathcal{E}^{de}_q$, \textit{amplitude-damping} channel $\mathcal{E}^{ad}_q$, and the \textit{mirrored} counterpart of the amplitude-damping, $\mathcal{E}_{q}^{mad}$.}
    \label{fig:ruido}
\end{figure}

In realistic scenarios, the experimental implementation of this protocol is often affected by sources of error, such as environmental noise or implementation imperfections, leading to the transference of a state $\rho^B_i$, with probability $p_i$, which is generally different from $\rho^a$ \cite{Oh2002,Knoll2014,Fortes2015,Im2021}, as illustrated in Fig. \ref{fig:ruido}. 
Consequently, two related and important questions arise: How to evaluate the performance of a non-ideal protocol? How to certify that a given teleportation protocol is genuinely quantum?

The average fidelity of teleportation $\braket{\bar{F}}$ is typically the chosen figure of merit for assessing protocol accuracy. It is defined as,
\begin{equation}\label{eq:avFid}
	\braket{\bar{F}} = \int \bar{F}(\rho^a)  \ d\psi,
\end{equation}
where $\rho^a=\ket{\psi}\!\bra{\psi}$, $\bar{F}(\rho^a) = \sum_i p_i F(\rho^a, \rho^B_i)$ is the score fidelity of the protocol and $F(\rho,\sigma) = \left(\Tr(\sqrt{\sqrt{\rho}\, \sigma \sqrt{\rho}})\right)^2$ is the Uhlman-Josza fidelity. The average is taken over all pure input states with $d\psi$ a normalized Haar measure: $\int d\psi =1$.

On the other hand, the literature widely states that a teleportation protocol is certified to be genuinely quantum as long as Alice can transmit the unknown initial state to Bob with greater accuracy compared to the best classical protocol, which stands for a \textit{measure-prepare} strategy (with no quantum resources and only one copy of the state to be teleported), achieving a maximum average fidelity of $\braket{\bar{F}}_{\cl} = 2/3$. 
Therefore, a teleportation protocol is quantum-certified only when the average fidelity of the protocol $\braket{\bar{F}}$ exceeds $2/3$, that is, $\braket{\bar{F}} > \braket{\bar{F}}_\cl = 2/3$~\cite{Massar1995,Pirandola2015}.

The choice of Uhlman-Josza fidelity as the main distinguishability measure between quantum states is crucial in this certification method. However, there are other equally valid options to consider, for instance, trace distance, Wootters distance, and Bures distance, among others. Although the possibility of using different distinguishability measures in teleportation was mentioned in the seminal work \cite{Popescu1994}, this concept has not been explored further. In this work, we aim to fill that gap by exploring the consequences of using different measures, specifically by analyzing the \textit{average distance of teleportation}:
\begin{equation}
    \label{eq:avg_distance_telep}
	\braket{\bar{D}} = \int \bar{D}(\rho^a)  \ d\psi,
\end{equation}
where $\bar{D}(\rho^a) = \sum_i p_i D(\rho^a, \rho^B_i)$ is the score distance of the protocol and $D(\rho,\sigma)$ is any distinguishability measure satisfying (i) positivity:  $D(\rho,\sigma) \geq 0$ with $D(\rho,\sigma) = 0$ iff $\rho=\sigma$, and (ii) unitary invariance: $D(\rho,\sigma) = D(U \rho U^\dag,U \sigma U^\dag)$. Property (i) defines a \textit{distance-like} distinguishability measure. In addition, we shall consider \textit{overlap-like} measures, such as fidelity or affinity, which reach their maximum value for $\rho=\sigma$; see Appendix \ref{app:distances} for further information on the distinguishability measures employed in this work.

First, in Sec. \ref{sec:measurementPrepare}, we obtain the optimal classical protocol for an arbitrary distance measure by minimizing the average distance of teleportation among all possible measurements and prepare protocols. Interestingly, this optimal protocol consists of two parts: one independent of the distance used, and the other dependent on it. Specifically, Alice is required to perform a projective measurement on her qubit and communicate the measurement result to Bob. Subsequently, Bob prepares a qubit with the same direction in the Bloch sphere as the observable corresponding to Alice's measurement result, in the same way as in Ref.~\cite{Massar1995} for the fidelity-based case. The purity of this output state is the only aspect that varies based on the chosen distance. Sec. \ref{sec:measurementPrepare}, therefore, stands for a generalization of the former optimal measurement and prepare strategy based on maximizing the average fidelity of teleportation~\cite{Massar1995}.

Second, we demonstrate that the conventional method of certifying the quantum nature of a particular protocol can lead to inconsistencies or be misleading when different distinguishability quantifiers are utilized. This can be found in Sec. \ref{sec:contradictions}, after reviewing the basics standard teleportation in Sec. \ref{sec:standard}. By analyzing several examples, we observe that employing various distance measures can result in contradictory conclusions. A particular teleportation protocol may be quantum-certified for a specific distance measure within a certain range of parameters, while it may not hold the same certification for a different distance measure within that same range of parameters. This behavior is particularly evident in a very common noisy experimental scenario~\cite{Urbanek2021}, where the resource state is a Werner state produced by the action of a depolarizing channel acting individually on the qubits of the Bell state. 
Importantly, this phenomenon, analyzed in Sec.~\ref{sec:unconditionallyoptimal}, is not exclusive to Werner states, as there exist other resource states that can yield conflicting conclusions regarding the quantum essence of the teleportation protocol. This is explored in detail in Section \ref{sec:otherexamples}.

The fact that different distance measures may lead to contradictory conclusions raises questions about the validity of the conventional approach. It also highlights the need for new certification methods that can provide a more reliable and comprehensive assessment of the success of quantum teleportation. 

\section{Measurement and prepare protocol}
\label{sec:measurementPrepare}

\subsection{Average distance of teleportation of measurement and prepare protocols}

Alice aims to teleport an unknown state of the qubit system $a$ to Bob using a classical strategy involving a \textit{measurement and prepare protocol} (MPP), assisted only by a classical communication channel. 
Let $\rho^a$ represent the initial pure state of system $a$ given by
\begin{equation}
	\rho^a = \frac{1}{2} \left(\Id + \vect \cdot \vecsigma \right),
\end{equation}
where $\vect$ is the Bloch vector of $\rho$ such that $|\vect| = 1$, and $\vecsigma =\left(\sigma_X, \sigma_Y,\sigma_Z\right)$ is the vector formed by the Pauli matrices.

The protocol starts with Alice taking a measurement over the qubit system $a$, characterized by a Positive Operator-Valued Measure (POVM) $\{E_i\}^n_{i=1}$, which can be parameterized as follows \cite{Vidal1999}:
\begin{equation}
	\label{eq:POVM_def}
	E_i = \frac{c^2_i}{2} \left(\Id + \vecs_i \cdot \vecsigma \right) ,
\end{equation}
with
\begin{equation}
	\label{eq:POVM_conditions}
	\sum^n_{i=1} c^2_i = 2 \ \text{and} \ \sum^n_{i=1} c^2_i \vecs_i= 0.
\end{equation}
In the next step, Alice classically communicates its measurement result to Bob, who then prepares the final state of the protocol according to a prefixed \textit{preparation strategy}: If Alice obtains the $i$th measurement result then Bob prepares the state
\begin{equation}
	\rho^B_i = \frac{1}{2} \left(\Id + \vecr_i \cdot \vecsigma \right),
\end{equation}
where $\vecr_i$ is the Bloch vector of $\rho^B_i$. 
The probability of obtaining this state is $p_i = \Tr\left(\rho E_i \right)$, which can be rewritten as
\begin{equation}
    \label{eq:pi_MPP}
	p_i = \frac{c_i^2}{2} \left(1 + \vect \cdot \vecs_i\right).
\end{equation}

The performance of the protocol can be assessed by calculating the average distance of teleportation, as given in Eq.~\eqref{eq:avg_distance_telep_fano}. 
For qubit systems, $D(\rho,\sigma)$ can be written as a function $d(\vec{r},\vec{s})$ with $\vec{r}$ and $\vec{s}$ being the Bloch vectors of states $\rho$ and $\sigma$, respectively. Refer to Appendix \ref{app:distances} for detailed expressions of $d(\cdot,\cdot)$ for some well-known distances. Consequently, for qubit input and output systems, the average distance of teleportation is
\begin{equation}
        \label{eq:avg_distance_telep_fano}
	\braket{\bar{d}\,}  = \frac{1}{4\pi}\int_S \ \bar{d}(\,\vect \,) \ d S, 
 %\left(\{\vecr_i\},\{\vecs_i\}\right)
\end{equation}
where $\bar{d}\left(\vect\right) = \sum^n_{i=1} p_i d(\vect,\vecr_i)$, being $\vect$ and $\vecr_i$ the corresponding Bloch vectors of $\rho^a$ and $\rho^B_i$, respectively. The integral is taken over the surface $S$ of the Bloch sphere.

We obtain that, for any measurement and prepare protocol, $\braket{\bar{d}\,}$ can be expressed as,
\begin{align}
	\label{eq:avg_distance_telep_MPP}
	\braket{\bar{d}}_{\mathrm{MPP}}  &= \sum^n_{i=1} \frac{c^2_i}{4}    \left[A_d(r_i) + \frac{\vecr_i \cdot \vecs_i}{r_i}   B_d(r_i) \right], 
\end{align}
where $A_d(r_i)= \int^1_{-1} g_d(z,r_i) dz $ and $B_d(r_i)= \int^1_{-1} g_d(z,r_i) z dz$, with $d(\vect,\vecr_i)=g_d(z,r_i)$, $r_i = \|\vec r_i\|$ and $z=\cos \theta$, being $\theta$ the angle between $\vec{t}$ and $\vecr_i$ (see Appendix~\ref{app:avg_distance_telep_MPP} for its explicit calculation).

In what follows, we will consider distinguishability measures that satisfy the property of faithfulness, that is, $\frac{\partial g_d(z,r_i)}{\partial z} \leq 0$ for $z \in [-1,1]$. Note that faithful overlap-like measures satisfy the opposite $\frac{\partial g_d(z,r_i)}{\partial z} \geq 0$; for more details, see Appendix \ref{app:distances}.

\subsection{Optimal minimal MPP}\label{sec:optMPP}

%\change{Let us consider now how to determine the optimal MPP protocol for transferring an unknown state of the system $a$ to $B$.} -- lo cambié a esto porque en realidad en esta sección no vemos cómo obtener el protocolo. 

Let us examine the optimal MPP protocol for transferring an unknown state of the system $a$ to $B$. We have to specify Alice's measurement and the preparation strategy by optimizing the considered figure of merit, which in this case corresponds to minimizing the average distance of teleportation, Eq.~\eqref{eq:avg_distance_telep_MPP}.

We have shown that for any distance measure $d(\cdot,\cdot)$, the optimal minimal Alice's measurement is an arbitrary standard von Neumann measurement, that is, anti-parallel unit Bloch vectors $\vecs_1$ and $\vecs_2$, and $c_1=c_2=1$, in Eq. \eqref{eq:POVM_def} (a detailed derivation can be found in Appendix~\ref{app:opt}). Given this measurement, the optimal preparation strategy is characterized by the preparation of the state
\begin{align}
    {\rho}^{\opt}_i=\frac{1}{2} \left(\Id + r^{\opt}\vecs_i \cdot \vecsigma \right),
\end{align}
when Alice communicates the $i$th result ($i=1,2$). Subsequently, as it is shown in Appendix \ref{app:opt}, the parameter $r^{\opt}$ is determined by the following minimization problem
\begin{equation}
	\label{eq:opt_norm_r}
	r^{\opt} = \argmin_{0 < r \leq 1} \int^1_{-1} g_d(z,r)(1+z)  dz.
\end{equation}
It is worth noting that the only step of the optimal minimal MPP protocol that depends on the specific distinguishability measure selected $d(\cdot,\cdot)$ is the preparation of the guessed state ${\rho}^{\opt}_i$, and this is exclusively through its purity $\mathcal{P}({\rho}^{\opt}_i)=[1+\left(r^{\opt}\right)^2]/2.$

Finally, for any given distance measure $d(\cdot,\cdot)$, we derive the optimal average distance as
\begin{equation}
	\label{eq:opt_dist}
	\braket{\bar{d}}_{\cl} = \frac{1}{2} \int^1_{-1} g_d(z,r^{\opt})(1+z)  dz.
\end{equation}
This quantity represents the classical threshold that any teleportation protocol has to surpass to be quantum-certified. Table~\ref{tab:table_of_d_min} shows the function $g_d$, the optimal norm $r^{\opt}$ and the minimal average distance $\braket{\bar{d}}_{\cl}$, for a set of known distinguishability measures, see Appendix \ref{app:distances}. 

Note that our method can also apply to fidelity or affinity, i.e. overlap-like measures, by solving the corresponding maximization problem instead of the minimization in Eq. \eqref{eq:opt_norm_r}.

\section{Contradictions in certifying quantum teleportation protocols} \label{sec:contradictions}

In the previous section, we established the optimal measurement and prepare protocol for different distinguishability measures $d(\cdot,\cdot)$, yielding classical thresholds $\braket{\bar{d}}_{\cl}$ (see Table~\ref{tab:table_of_d_min}). This enables us to generalize the certification method in teleportation beyond the usual quantum fidelity. Specifically, a teleportation protocol is \textit{quantum-certified} if its average teleportation distance fails the classical threshold ($\avgd < \avgd_\cl$); otherwise, it is \textit{classical-certified}.

\subsection{Noisy standard teleportation protocol}\label{sec:standard}

Let us first recall the basis of noisy standard teleportation \cite{Horodecki1996}. 
This particular teleportation protocol works with a global quantum system constituted by three qubits $a$, $A$ and $B$, operated by the sender (Alice) and the receiver (Bob). Alice has to transfer the unknown state occupying the qubit system $a$ to system $B$, usually referred to as Bob's qubit. The standard quantum teleportation protocol takes advantage of the potential correlations between qubits $A$ and $B$, described by a joint state $\rho^{AB}$ known as \textit{resource state}. By employing the Fano form of an arbitrary two-qubit state, we can write, 
\begin{small}
    \begin{align}\label{eq:ResourceStateAB}
\rho^{AB}\!=\!\frac{1}{4}\!\left(\mathbb{1}\!\otimes\!\mathbb{1} + \vec{r}^A\!\cdot\!\vec{\sigma}\! \otimes\!\mathbb{1} + \mathbb{1}\!\otimes\!\vec{r}^B\!\cdot\!\vec{\sigma}\!+\!\sum_{ij=1}^3 r_{ij}\sigma_i\!\otimes\!\sigma_j \right)\!,
\end{align}
\end{small}

\noindent where $\mathbb{1}$ is the identity operator in the space of the corresponding qubit system, $\vec{r}^A$ and $\vec{r}^B$ are the Bloch vectors of the reduced states $\rho^{A} = \Tr_{B} \rho^{AB}$ and $\rho^{B} = \Tr_{A} \rho^{AB}$:
\begin{equation}
	\rho^{A} = \frac{1}{2}\left(\mathbb{1}+\vec{r}^{A} \cdot\vec{\sigma}\right) \ \text{and} \ \rho^{B} = \frac{1}{2}\left(\mathbb{1}+\vec{r}^{B} \cdot\sigma\right),
\end{equation}
and $r_{ij}$ are the elements of the correlation matrix $\mathbb{r} = \Tr\left(\rho^{AB} \,\vec{\sigma} \otimes \vec{\sigma} \right).$

\onecolumngrid

\begin{table}[h!]
	\centering
	\begin{tabular}{lccc}
		\toprule
		Distinguishability measure & $g_d$ & $r^{\opt}$ & $\braket{\bar{d}}_{\cl}$  \\ \hline 
  Trace distance~\cite{Fuchs1996,Nielsen2010} & $\frac{\sqrt{1-2rz+r^2}}{2}$ & $\frac{2 \sqrt{10}-5}{3} \approx 0.441$ & $\frac{8(11 - 2 \sqrt 10)}{81} \approx 0.461 $ \\ [0.2cm]
     Wootters distance~\cite{Wootters1981} & $\arccos\sqrt{\frac{1}{2}(1+rz)}$ & 1 & $\approx 0.589$ \\ [0.2cm]
  Bures distance~\cite{Uhlmann1976,Jozsa1994} & $\sqrt{2(1-\sqrt{\frac{1+rz}{2}})}$ & 1 & $\frac{128 \sqrt{2}}{315} \approx 0.574$ \\ [0.2cm]
		Affinity~\cite{Luo2004} & $A(z,r)=\frac{1+rz+\sqrt{1-r^2}}{\sqrt{2}(\sqrt{1+r}+\sqrt{1-r})}$ & 3/5 & $\frac{\sqrt{5}}{3}\approx 0.745$ \\ [0.2cm]
		Hellinger distance~\cite{Luo2004} &$2-2A(z,r)$ & 3/5 & $2 - \frac{2 \sqrt{5}}{3}\approx 0.509 $ \\ [0.2cm]
            \begin{tabular}{l}
		  Quantum Jensen-Shannon \\
		divergence~\cite{Majtey2005} 
		\end{tabular} &
		  \begin{tabular}{c}
		   $H_2\left(\frac{\sqrt{1+rz+r^2}}{2}\right)-H_2(r)$  \\
		     with $H_2(r)=-\frac{1+r}{2} \log_2\frac{1+r}{2} - \frac{1-r}{2} \log_2\frac{1-r}{2}$
		\end{tabular} 
            & $\approx 0.426$ & $\approx 0.282$ \\  [.5cm]
		Transmission distance~\cite{Briët2009,Bussandri2023}  & $\sqrt{H_2\left(\frac{\sqrt{1+rz+r^2}}{2}\right)-H_2(r)}$ & $\approx 0.528$ & $\approx 0.524$ \\
		\midrule
		Fidelity~\cite{Uhlmann1976,Jozsa1994} & $\frac{1}{2}(1+rz)$ & $1$ & $2/3$ \\
		\bottomrule
	\end{tabular}
\caption{For each distinguishability measure listed in the first column, we provide the corresponding form of the function $g_d$, the norm $r^{\opt}$ of the optimal state, and the classical threshold $\braket{\bar{d}}_{\cl}$.}
\label{tab:table_of_d_min}
\end{table}
\twocolumngrid
 
Once the resource state is prepared, Alice measures a projection onto the Bell basis, defined by the four Bell states
$\{\ket{\Phi_i}\}_{i=1}^4$, being:
\begin{align}
  \ket{\Phi_{1}}&\!=\!\frac{1}{\sqrt{2}}\!\left(\ket{00}\!+\!\ket{11} \right), &  \ket{\Phi_{3}}&\!=\! \frac{1}{\sqrt{2}}\!\left(\ket{01}\!+\!\ket{10} \right), \label{eq:Bellstates} \\ 
\ket{\Phi_{2}}&\!=\! \frac{1}{\sqrt{2}}\!\left(\ket{00}\!-\!\ket{11} \right), & \ket{\Phi_{4}}&\!=\! \frac{1}{\sqrt{2}}\!\left(\ket{01}\!-\!\ket{10} \right).  \label{eq:Bellstates1} 
\end{align}

The corresponding measurement operators are therefore $\{E^{aA}_i=\ket{\Phi_i}\!\bra{\Phi_i}\}_{i=1}^4$. 
The Fano form of these projectors, see Eq. \eqref{eq:ResourceStateAB}, is given by null Bloch vectors for the marginal states and the following correlation matrices:
\begin{align}\label{eq:BellstatesCorrMat}
	\mathbb{w}_1&=\text{diag}(1,-1,1), &  \mathbb{w}_3&=\text{diag}(1,1,-1),\\
	\mathbb{w}_2&=\text{diag}(-1,1,1), & \mathbb{w}_4&=\text{diag}(-1,-1,-1). \label{eq:BellstatesCorrMat1}
\end{align}
If Alice obtains the $i$th measurement outcome, the resulting state occupying system $B$ is~\cite{Toscano2023}
\begin{align}\label{eq:CondState}
    \rho^{B|i} &= \frac{1}{2} \left(\Id + \vect^{\,B|i} \cdot \vecsigma \right), \ \text{where} \\
    \vect^{\,B|i}&=\frac{\vec{r}^B+(\mathbb{w}_i \mathbb{r})^\intercal \vec{t}}{4 p_i}, \text{ with } p_i=\frac{1+ \vec{t}\cdot (\mathbb{w}_i\vec{r}^A)}{4}. \label{eq:conditionalVect}
\end{align} 
Finally, Alice communicates its measurement result (let us assume the $i$th outcome) to Bob who applies a unitary operation $U_i$ over the qubit system $B$ in order to recreate the original state $\rho^a$:
\begin{equation}
\label{eq:outputState}
	\rho^{B}_i =U_i \rho^{B|i} U_i^\dagger=\frac{1}{2}(\mathbb{1}+\vec{t}_i \cdot \vec{\sigma}), \ \text{and } \vec{t}_i=\mR_i\vec{t}^{\,B|i}, 
\end{equation}
where $\mR_i$ is the unique rotation in the Bloch sphere defined by $U_i$.

\subsection{Unconditionally optimal noisy standard protocol} \label{sec:unconditionallyoptimal}

\begin{figure*}[htb]\centering
	\includegraphics[width=\textwidth]{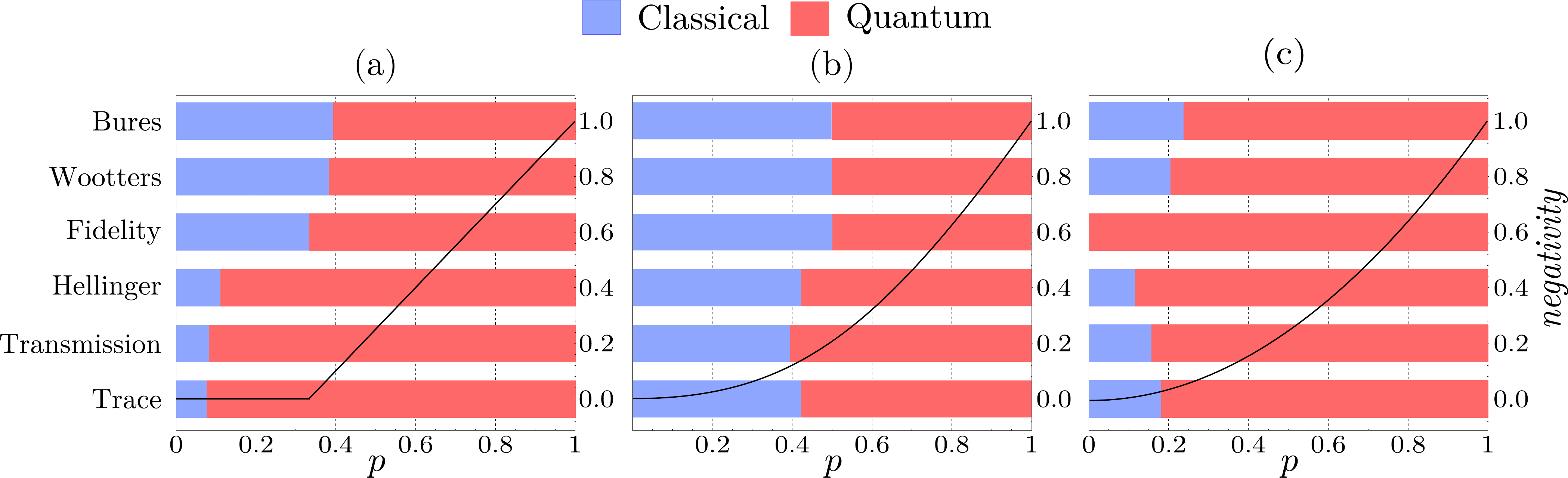}  
	\caption{How different distinguishability measures certify the protocol as either classical (blue) or quantum (red) in terms of the parameter $p$ of three different resource states: (a) the Werner state in Eq.~\eqref{eq:Werner_state}; (b) the state in Eq.~\eqref{eq:calvacanti_resource}; and (c) the state in Eq. \eqref{eq:twoADCresource}. The solid line depicts the normalized negativity (entanglement) of the resource state in terms of the parameter $p$.}
 \label{fig:barras}
\end{figure*}

For a given quantum resource state $\rho^{AB}$, the average teleportation distance corresponding to the noisy standard protocol can be calculated by introducing the conditional states $\rho^{B}_i$ in Eq.~\eqref{eq:avg_distance_telep}. This quantity therefore depends on Bob's rotations $\{\mR_i\}_i$, namely,  $\avgd = \avgd\left(\left\{ \mR_i \right\}\right)$.

We define a standard protocol as \textit{conditionally optimal} if, for a particular distinguishability measure $d(\cdot,\cdot)$, the rotations performed by Bob minimize the average teleportation distance. Similarly, a standard protocol is \textit{unconditionally optimal} if the rotations executed by Bob minimizing the average teleportation distance are the same for any distinguishability measure.

When a qubit of a Bell resource state $\ket{\Phi_k}\bra{\Phi_k}$ is affected by a depolarizing noise $\mathcal{E}^{de}_q(\rho)=q\rho+\frac{(1-q)}{2}\mathbb{1}$, the resource becomes a Werner state
\begin{equation}
\label{eq:Werner_state}
\rho^{AB}_\mathrm{W} = \mathcal{E}^{de}_{q_A}\!\otimes\mathcal{E}^{de}_{q_B}(\ket{\Phi_k}\bra{\Phi_k})\! =\! p\ket{\Phi_k}\bra{\Phi_k} + \frac{(1-p)}{4}\mathbb{1}
\end{equation}
being the parameter $p=q_{A}q_{B}$, see Appendix~\ref{app:localnoises}. In this case, the standard protocol becomes unconditionally optimal (as detailed in Appendix~\ref{app:Werneruosp}). The optimal Bob rotations are
\begin{equation}
    \label{eq:opt_rot_werner}
    \mR^{\opt}_i =\mathbb{w}_{i}\mathbb{w}_k,
\end{equation}
where $\mw_i$ are the Bell state correlation matrices, Eqs. \eqref{eq:BellstatesCorrMat}-\eqref{eq:BellstatesCorrMat1}, and $\mathbb{r} _W = p \mathbb{w}_{k}$ is the correlation matrix of the Werner state. However, this unconditionality does not guarantee that the certification method remains independent of the chosen distinguishability measure. In fact, we observe contradictions among all the distinguishability measures listed in Table~\ref{tab:table_of_d_min}.
Figure \ref{fig:barras}.(a) depicts how various distinguishability measures classify the protocol as either classical ($\avgd \geq \avgd_\cl $) or quantum ($\avgd < \avgd_\cl$) based on the parameter $p$ (for fidelity and affinity, both inequalities are reversed).
The Bures distance enforces the strictest criterion, giving the smallest interval of values of $p$ for which the protocol is certified as quantum. In contrast, the trace distance proves to be the least restrictive. 
Notably, the maximum discrepancy interval (i.e. the greater difference between two transition points from classical to quantum certification) is $\Delta p \approx 0.317$.

We encounter, additionally, a counterintuitive result for the trace distance, transmission distance, and Hellinger distance: the standard protocol is certified as quantum even for Werner separable states ($p \leq 1/3$).

See Appendix~\ref{app:averDistExamples} for a complete description of the average distance of teleportation cases considered in this example and the negativity (entanglement) of the resource state.

\subsection{Other resource states from amplitude-damping channels} \label{sec:otherexamples}
\subsubsection{Example 1}
%\diego{\subsection{Conditionally optimal noisy standard protocol}}
\label{sec:cavalcantiprot}

Consider the standard protocol with a Bell state resource $\ket{\Phi_1}$ affected by two local noises (as illustrated in Fig.\ref{fig:ruido}) described by an \textit{amplitude-damping} channel $\mathcal{E}^{ad}_{q}(\rho)=\sum_{i=1}^2 A_i \rho A_i^{\dagger}$, where~\cite{Bengtsson2006}
\begin{align}\label{eq:kraussopADC}
A_1 &= \begin{bmatrix}
1 & 0 \\
0 & \sqrt{q}
\end{bmatrix} & A_2 &= \begin{bmatrix}
0 & \sqrt{1-q} \\
0 & 0
\end{bmatrix},
\end{align}
and its mirrored counterpart~\cite{Knoll2018}, $\mathcal{E}^{mad}_{q}(\rho)=\sum_{i=1}^2 B_i \rho B
_i^{\dagger}$, being
\begin{align}\label{eq:MirroredKraussopADC}
B_1 &= \begin{bmatrix}
1 & 0 \\
0 & \sqrt{q}
\end{bmatrix} & B_2 &= \begin{bmatrix}
0 & 0 \\
\sqrt{1-q} & 0
\end{bmatrix}.
\end{align}

%\MB{DEFINIR LOS OPERADORES DE KRAUS DEL OTRO CANAL Y LLAMARLO AD-LIKE, MOSTRAR EN EL APENDICE QUE ESTOS DOS CANALES LLEVAN AL EJEMPLO ESTE.}

The resulting resource state is given by
\begin{align}
\rho^{AB}_{ad} &= \mathcal{E}^{ad}_{p}\otimes\mathcal{E}^{mad}_{p} (\ket{\Phi_1}\!\bra{\Phi_1}) \nonumber\\
&=p\ket{\Phi_1}\bra{\Phi_1} + (1-p)\ket{01}\bra{01}, \label{eq:calvacanti_resource}
\end{align}
where $p \in [0,1]$. 
This state has also been studied in Ref.~\cite{Cavalcanti2017}, whereas the local noises have been experimentally implemented in \cite{Knoll2018}.
See Appendix \ref{app:localnoises} for more details on this example.

In this case, the optimal Bob rotations maximizing the average fidelity are
\begin{equation}
\label{eq:opt_Bob_rotations_calvacanti}
    \mR^{\opt}_i = \begin{cases}
        \mathbb{w}_{i}\mathbb{w}_4 \ \text{ if } \ p<1/3,\\
        \mathbb{w}_{i}\mathbb{w}_1 \ \text{ if } \ p\geq1/3,
    \end{cases}
\end{equation}
where $\mathbb{w}_i$ are defined in Eqs. \eqref{eq:BellstatesCorrMat}-\eqref{eq:BellstatesCorrMat1} (consult Appendix~\ref{app:BobRotFid} for a comprehensive derivation of Eq. \eqref{eq:opt_Bob_rotations_calvacanti}).

This protocol is designed to maximize fidelity and not necessarily optimize other distinguishability measures; Moreover, the curves corresponding to the average distance of teleportation included in Appendix \ref{app:averDistExamples} imply that this protocol is conditionally optimal. 
Nonetheless, it is interesting to explore whether this protocol is classified as classical or quantum, depending on the parameter $p$, if other distinguishability measures are used. Figure \ref{fig:barras}.(b) illustrates this point, revealing contradictory and counterintuitive findings.

% For example, the trace distance quantum-certified this protocol for all $p\in[0,1]$, a result that could be supported by the fact that this particular resource state is entangled for any value of the parameter $p$; but this is observed solely for this measure. For the other considered distinguishability measures, always exit a range in which this protocol is certified as classical. Remarkably, the same hierarchy observed in the Werner case persists, in the sense that Bures distance imposes the most stringent criterion and the trace distance proves to be the least restrictive.

In this case, although Bures and Wootters distances give rise to similar certifications as fidelity, there always exists a range of $p$-values implying discrepancy: The transition from classical to quantum certifications takes place at $p\approx 0.49845$ for Bures distance, while for Wootters distance $p\approx 0.49617$. These transition points differ from $p=1/2$ which corresponds to fidelity (small improvements in the average fidelity of teleportation that change the nature of the certification have been reported in Ref. \cite{Badziag2000}).
%While these variations may seem minor in scale, they are significant as the nature of a protocol certification fundamentally changes, moving from classical to quantum or vice versa. 
% \diego{Note that, analogously, small improvements of the average fidelity of teleportation changing the nature of the protocol certification has been reported for a different noise model in Ref. \cite{Badziag2000}.}

Trace, transmission, and Hellinger distances produce bigger conflicting ranges, being the transmission distance the least restrictive ($p\approx 0.39348$). Notably, the most stringent criterion for this protocol corresponds to fidelity.

We have included in Appendix~\ref{app:averDistExamples}, a complete description of the average distance of teleportation cases considered in this example and the negativity of the resource state.

\subsubsection{Example 2} \label{sec:twoADC}

In the previous example, we analyzed the standard teleportation protocol where qubits $A$ and $B$ were affected by the amplitude-damping channel and its mirrored version, respectively. Now, let us consider a slight modification in which the resource state qubits are equally subjected to two amplitude-damping channels \cite{Knoll2014}:
\begin{align}\label{eq:twoADCresource}
\Tilde{\rho}^{AB}_{ad} &= \mathcal{E}^{ad}_{p}\otimes\mathcal{E}^{ad}_{p} (\ket{\Phi_1}\!\bra{\Phi_1}),
\end{align}
with $\mathcal{E}^{ad}_{p}$ given by Eq. \eqref{eq:kraussopADC}. This minor modification changes the corresponding correlation matrix of the resource state, leading to a change in the optimal Bob rotations to:
\begin{equation}
\label{eq:bob_opt_twoADC}
    \mR^{\opt}_i = \mathbb{w}_{i}\mathbb{w}_1. 
\end{equation}
Full details on the previous calculations can be found in Appendices \ref{app:localnoises} and \ref{app:BobRotFid}.

Figure \ref{fig:barras}.(c) shows how the average teleportation distances certify the protocol as classical or quantum. Once again, all distance measures lead to contradictory classifications.

The average fidelity leads to a quantum certification for all $p$, as was already observed in Ref. \cite{Knoll2014}. Remarkably, all other considered distance measures have a range of $p$ in which the certification is classical, even Wootters and Bures distances which are functions of fidelity.

On the other hand, there is no observed same ordering concerning stringency among the distance measures of the other noise models, with the Bures distance and fidelity being the most and least stringent, respectively, in this case.

%\change{In Ref. \cite{Knoll2014} was examined theoretically and experimentally, among different noise models, the average fidelity of teleportation for this particular teleportation setup: It should be noted that the certification results quantum for all values of $p$.}

\section{Concluding remarks}

In the domain of quantum information, quantum teleportation plays a key role as a fundamental protocol. The theoretical ideal guarantees perfect state transfer that cannot be emulated with any classical protocol, but practical implementations face errors arising from the environment. Consequently, evaluating and certifying the performance of a protocol becomes essential. Traditionally, average fidelity has served as the usual benchmark, requiring exceeding the $2/3$ threshold for quantum certification.

Our contribution delves into an unexplored investigation by adopting alternative measures of distinguishability. Although some of them have a clear physical meaning in other contexts, such as quantum metrology, in quantum teleportation it is still not clear in physical terms why a distinguishability measure has to prevail over another. In light of this, we introduced and employed the concept of average teleportation distance to assess the protocol performance.

Initially, we derived the optimal classical measurement and preparation protocol for any chosen distinguishability measure (Sec.~\ref{sec:measurementPrepare}). Interestingly, the optimality of the classical strategy is contingent on the employed distance only in the purity of the prepared state. 

Subsequently, we unveiled a critical issue: the conventional certification method can lead to contradictory interpretations when different distinguishability measures are applied (Sec. \ref{sec:contradictions}). 
This phenomenon was observed through examinations of different noise models affecting the resource state of the standard teleportation protocol.
Specifically, we analyzed three cases: depolarizing channels, which result in a Werner state as the resource for the protocol (refer to Eq.\eqref{eq:Werner_state}); amplitude damping channel and its mirrored counterpart, leading to the resource state described in Eq.\eqref{eq:calvacanti_resource}; and two equal amplitude damping channels, resulting in the resource state given by Eq.~\eqref{eq:twoADCresource}.

This study advocates for a reevaluation of established methodologies and underscores the pressing necessity for more comprehensive and consistent certification strategies.

\begin{acknowledgments}
DGB was supported by the Spanish Ministry of Science and Innovation (MCIN) with funding from the European Union Next Generation EU (PRTRC17.I1) and the Department of Education of Castilla y León (JCyL) through the QCAYLE project, as well as MCIN projects PID2020-113406GB-I00 and RED2022-134301-T. GMB acknowledges financial support from project PIBAA 0718 funded by Consejo Nacional de Investigaciones Cient\'ificas y T\'ecnicas CONICET (Argentina). FT acknowledges financial support from the Brazilian agency INCT-Informação Quântica, CAPES, and CNPq.

\end{acknowledgments}

\newpage
\onecolumngrid
\appendix

\section{Distinguishability measures between quantum states}
\label{app:distances}

A distance-like distinguishability measure between two quantum states $\rho$ and $\sigma$ is a functional $D(\rho,\sigma)$ satisfying (i) positivity:  $D(\rho,\sigma) \geq 0$ with $D(\rho,\sigma) = 0$ iff $\rho=\sigma$, and (ii) unitary invariance: $D(\rho,\sigma) = D(U \rho U^\dag,U \sigma U^\dag)$. Table~\ref{tab:distances} provides a summary of the expressions for distinguishability measures applied to qubit systems, used on the average teleportation distance. Note that we also take into account overlap-like measures, such as Fidelity and Affinity, which are positive quantities that satisfy unitary invariance and attain their maximum value when $\rho=\sigma$.

\begin{table}[h!]
	\centering
	\begin{tabular}{lcc}
		\toprule
		Distinguishability measure & General ($D(\rho,\sigma)$) & Qubits ($d(\vec{r},\vec{s})$) \\ \hline
		Trace distance~\cite{Fuchs1996,Nielsen2010} & $D_{\Tr}(\rho,\sigma)=\frac{1}{2} \Tr \left|\rho-\sigma\right|$, with  $|O|= \sqrt{O^\dagger O}$ & $d_{\Tr}(\vec{r},\vec{s})=\frac{1}{2}\sqrt{(\vec{r}-\vec{s})^2}$ \\[0.1cm]
            Fidelity~\cite{Uhlmann1976,Jozsa1994} & $F(\rho,\sigma)=\Tr\left[\sqrt{\sqrt{\rho}\sigma\sqrt{\rho}}\right]^2$ & $f(\vec{r},\vec{s})=\frac{1}{2}\left(1+\vec{r}\cdot\vec{s}+\sqrt{1-\vec{r}^2}\sqrt{1-\vec{s}^2}\right)$ \\[0.1cm]
    	Wootters distance~\cite{Wootters1981} & $D_{\mathrm{W}}(\rho,\sigma) = \arccos\sqrt{F(\rho,\sigma)}$ & $d_{\mathrm{W}}(\vec{r},\vec{s})=\arccos\sqrt{F(\rho,\sigma)}$  \\[0.1cm]
		Bures distance~\cite{Uhlmann1976,Jozsa1994} & $D_{\mathrm{B}}(\rho,\sigma) = \sqrt{2(1-\sqrt{F(\rho,\sigma)})}$ &   $d_{\mathrm{B}}(\vec{r},\vec{s}) = \sqrt{2(1-\sqrt{f(\vec{r},\vec{s}})}$ \\[0.1cm]
		Affinity~\cite{Luo2004} & $A(\rho,\sigma)=\Tr \sqrt{\rho}\sqrt{\sigma}$ & $a(\vec{r},\vec{s})=\frac{\vec{r}\cdot\vec{s}+(1+\sqrt{1-\vec{r}^2})(1+\sqrt{1-\vec{s}^2})}{(\sqrt{1+|\vec{r}|}+\sqrt{1-|\vec{r}|})(\sqrt{1+|\vec{s}|}+\sqrt{1-|\vec{s}|})}$ \\[0.1cm]
		Hellinger distance~\cite{Luo2004} &$D_{\mathrm{H}}(\rho,\sigma) =2-2A(\rho,\sigma)$ & $d_{\mathrm{H}}(\vec{r},\vec{s}) =2-2a(\vec{r},\vec{s})$ \\[0.1cm]
  \hspace{-.189cm}
\begin{tabular}{l}
\noindent Quantum Jensen-Shannon  \\
\noindent divergence~\cite{Majtey2005} 
\end{tabular} & 
            \begin{tabular}{c}
            $D_{\mathrm{QJSD}}(\rho,\sigma) = S\left(\frac{\rho+\sigma}{2}\right)-\frac{1}{2}S(\rho)-\frac{1}{2}S(\sigma)$,   \\
            with $ S(\tau)= -\Tr\left[\tau \log_2 \tau\right]$     
            \end{tabular}
            & 
            \begin{tabular}{c}
            $d_{\mathrm{QJSD}}(\vec{r},\vec{s}) = s\left(\frac{\vec{r}+\vec{s}}{2}\right)-\frac{1}{2}s(\vec{r})-\frac{1}{2}s(\vec{s})$,   \\
            with $s(t) = - \frac{1+t}{2} \log_2\frac{1+t}{2}- \frac{1-t}{2} \log_2\frac{1-t}{2}$     
            \end{tabular}
            \\[0.1cm]
		Transmission distance~\cite{Briët2009,Bussandri2023} & $D_{\mathrm{Ts}}(\rho,\sigma)=\sqrt{D_{\mathrm{QJSD}}(\rho,\sigma)}$ & $d_{\mathrm{Ts}} = \sqrt{d_{\mathrm{QJSD}}(\vec{r},\vec{s})}$\\
  \bottomrule
	\end{tabular}
\label{tab:distances}
\caption{Expressions for distinguishability measures applied to qubit systems $\rho=\frac{1}{2}(\mathbb{1}+\vec{r}\cdot\vec{\sigma})$ and $\sigma=\frac{1}{2}(\mathbb{1}+\vec{s}\cdot\vec{\sigma})$.}
\end{table}

For qubit systems, we express the distinguishability measure between $\rho=\frac{1}{2}(\mathbb{1}+\vec{r}\cdot\vec{\sigma})$ and $\sigma=\frac{1}{2}(\mathbb{1}+\vec{s}\cdot\vec{\sigma})$ as a function of the corresponding Bloch vectors: $D(\rho,\sigma)=d(\vec{r},\vec{s})$. Having in mind the unitary invariance, it follows that, if $|\vec{r}|=1$ and $|\vec{s}|=s$, $d(\vec{r},\vec{s})=g_d(z,s)$ with $z=\cos{\theta}$ being $\theta$ the angle between $\vec{r}$ and $\vec{s}$. Finally, the natural property of faithfulness for distance-like (overlap-like) measures $d(\cdot,\cdot)$ is $\frac{\partial g_d(z, r_i)}{\partial z} \leq(\geq) \ 0$ for $z \in [-1,1]$.

\section{Proofs of results given in Section \eqref{sec:measurementPrepare}}
\label{app:Proofs of MPP}
\subsection{Average distance of teleportation for a measurement and prepare protocol}
\label{app:avg_distance_telep_MPP}

Let us obtain the average distance of teleportation (Eq. \eqref{eq:avg_distance_telep_fano} of the main article, for qubit input and output systems) corresponding to a measurement and prepare protocol, characterized by Eqs. \eqref{eq:POVM_def} and \eqref{eq:POVM_conditions} of the main article. 

By inserting $p_i = \frac{c_i^2}{2} \left(1 + \vect \cdot \vecs_i\right)$ in 
$\braket{\bar{d}\,}  = \frac{1}{4\pi}\int_S \ \sum_i p_i d(\vec{t},\vec{r}_i) \ d S$ the average distance of teleportation is: 
\begin{equation}\label{eq:MPPAppendix1}
    \braket{\bar{d}}  = \frac{1}{4\pi}  \sum^n_{i=1} \frac{c_i^2}{2}  \left[\int_S    d(\vect,\vecr_i) \  dS+\int_S   \vect \cdot \vecs_i\, d(\vect,\vecr_i) \  dS\right].
\end{equation} 
As $d(\cdot,\cdot)$ is invariant under rotations, $d(\vec{t},\vec{r}_i)$ is just a function of $r_i$ and the angle $\theta$ between $\vec{t}$ and $\vec{r}_i$. For convenience, we shall denote $d(\vec{t},\vec{r}_i)=g_d(z,r_i)$, where $z=\cos \theta$. Besides, due to $\braket{\bar{d}}$ is invariant under $\vec{t} \to R \vec{t}$, $\vec{s}_i \to R \vec{s}_i$ and $\vec{r}_i \to R \vec{r}_i$, with $R$ an arbitrary rotation,  we can take the following coordinate system: $\vecr_i = r_i \hat{z}$, $\vecs_i = s_i \left( \sin\theta'_i \hat{x}+ \cos\theta'_i \hat{z} \right)$ with $\theta'_i$ the angle between $\vecs_i$ and $\vecr_i$, and $\vect = \cos\phi \sin\theta \hat{x} + \sin\phi \sin\theta \hat{y} + \cos\theta \hat{z} $.

Returning to Eq. \eqref{eq:MPPAppendix1}, we have that 
$$\frac{1}{2\pi} \int_S    d(\vect,\vecr_i) \  dS=\int^1_{-1}  g_d(z,r_i) \  dz = A_d(r_i),$$
where we have employed $z=\cos\theta$. The second term is given by, 
\begin{align*}
    \int_S \vect \cdot \vecs_i d(\vect,\vecr_i)  dS  = \int_0^{2\pi}  \int_0^\pi   s_i (\cos\phi \sin \theta \sin \theta'_i+\cos\theta\cos\theta'_i) \times  g_d(\cos \theta,r_i)   \sin \theta  d\theta  d\phi=2\pi\frac{\vecr_i \cdot \vecs_i}{r_i}B_d(r_i)
\end{align*}

\noindent being $B_d(r_i)=\int^1_{-1}  g_d(z,r_i) z \  dz$. Thus, we have seen that Eq. \eqref{eq:MPPAppendix1} is equivalent to the Eq. \eqref{eq:avg_distance_telep_MPP} of the main article:
$$\braket{\bar{d}}_{\mathrm{MPP}}  = \sum^n_{i=1} \frac{c^2_i}{4}    \left[A_d(r_i) + \frac{\vecr_i \cdot \vecs_i}{r_i}   B_d(r_i) \right].$$

\subsection{Optimal guessed state} \label{app:opt}

We aim to demonstrate that $B_d(r_i) = \int_{-1}^{1} g_d(z, r_i)z dz$ is negative (positive) for any distance-like (overlap-like) faithful distinguishability measure $d(\cdot,\cdot)$. Applying integration by parts, we get: 
\begin{align*}
    B_d(r_i)= \frac{1}{2} \int_{-1}^{1} \frac{\partial g_d(z,r_i)}{\partial z} dz  -\frac{1}{2} \int^{-1}_1 \frac{\partial g_d(z,r_i)}{\partial z} z^2 dz =\frac{1}{2}\int_{-1}^{1} \frac{\partial g_d(z,r_i)}{\partial z}  (1-z^2) dz.
\end{align*}
Since $\frac{\partial g_d(z, r_i)}{\partial z} \leq (\geq) \ 0$ and $1-z^2 \geq 0$ for $z \in [-1,1]$ (faithfulness property), it follows that $B_d(r_i) \leq (\geq) \ 0$. 
%\change{Note that for overlap-like measures, the faithful property is $\frac{\partial g_d(z, r_i)}{\partial z} \geq 0$ and $B_d(r_i)$ is positive. For the purpose of this section, we will focus on distance-like measures. The extension to overlap-like cases is straightforward.}

Now, for the objective function $\braket{\bar{d}}\!(\{r_i,s_i,\theta'_i\})=\sum^n_{i=1} \frac{c^2_i}{4}    \left[A_d(r_i) + s_i \cos\theta'_i   B_d(r_i) \right]$, we have to solve:
\begin{mini!}|l|[2]
	{\{r_i,s_i,\theta'_i\}}{\braket{\bar{d}}(\{r_i,s_i,\theta'_i\}),}
	{}{}
	\addConstraint{0 < r_i,s_i \leq 1 \ \text{and} \ 0\leq\theta'_i\leq2\pi }. 
\end{mini!}

Let us continue assuming the distance-like case. Defining $f_d(r_i,s_i,\theta'_i)=A_d(r_i) + s_i \cos\theta'_i   B_d(r_i)$ and due to the negative value of $B_d(r_i)$, we have that for any $r_i$, $s_i$, and $\theta'_i$ satisfying the constraints, it holds $f_d(r_i,s_i,\theta'_i) \geq f_d(r_i,s_i,0) \geq f_d(r_i,1,0)\geq f_d(r^\opt,1,0)$ where 
\begin{align}
    r^\opt &= \argmin_{0 < r \leq 1}   \left[A_d(r) +  B_d(r) \right],  \\
    &=  \argmin_{0 < r \leq 1} \int^1_{-1} g_d(z,r)(1+z)  dz.
\end{align}
Therefore, it follows
\begin{equation}
	\min_{\{r_i,s_i,\theta_i'\}}\braket{\bar{d}}\!(\{r_i,s_i,\theta'_i\})= \braket{\bar{d}}_{\cl} = \frac{1}{2} \int^1_{-1} g_d(z,r^{\opt})(1+z)  dz.
\end{equation}

Note that for the case of overlap-like measures, it is necessary to solve:
\begin{maxi!}|l|[2]
	{\{r_i,s_i,\theta'_i\}}{\braket{\bar{d}}(\{r_i,s_i,\theta'_i\}),}
	{}{}
	\addConstraint{0 < r_i,s_i \leq 1 \ \text{and} \ 0\leq\theta'_i\leq2\pi }.
\end{maxi!}
This optimization is completely analogous to the previous case of distance-like distinguishability measures.

As observed, the optimal strategy is independent of the number $n$ of elements in Alice's measurement. Consequently, the optimal \textit{minimal} strategy entails performing an arbitrary projective measurement with $n=2$, employing anti-parallel Bloch vectors $\vecs_1=-\vecs_2$ (to fulfill the POVM conditions) with unit norm ($s_1 = s_2 =1$). Subsequently, Bob has to prepare states with Bloch vectors $\vecr_i$ aligned with $\vecs_i$, and with equal norms given by $r_i = r^\opt$. Thus, it is clear that Bob's preparation strategy depends on the choice of the distinguishability measure $d(\cdot,\cdot)$.

\section{Resource state as a Bell state affected by local noises}
\label{app:localnoises}

Let us recall that the action of two quantum channels $\mathcal{E}$ and $\mathcal{F}$ with affine decomposition $(A_{\mathcal{E}},\vec{b}_{\mathcal{E}})$ and $(A_{\mathcal{F}},\vec{b}_{\mathcal{F}})$, respectively, over a Bell state $\left|\Phi_k\right>\!\left<\Phi_k\right|$ results into the resource state~\cite{Toscano2023},
\begin{align}\label{eq:noiseschannels}
     \mathcal{E} \otimes \mathcal{F}\left(\left|\Phi_k\right>\!\left<\Phi_k\right|\right) = \frac{1}{4}\left[\mathbb{1}_4+\vec{b}_{\mathcal{E}} \cdot \vec{\sigma} \otimes \mathbb{1}+\mathbb{1} \otimes \vec{b}_{\mathcal{F}} \cdot \vec{\sigma}+\sum_{\alpha \beta = 1}^3\left(b_{\mathcal{E}}^\alpha b_{\mathcal{F}}^\beta+\sum_{i=1}^3\mathbb{w}_k^{ii} A_{\mathcal{E}}^{\alpha i} A_{\mathcal{F}}^{\beta i}\right) \sigma_\alpha \otimes \sigma_\beta\right].
\end{align}

The depolarizing channel $\mathcal{E}_q^{de}$ acts according to: 
\begin{align}
    \vec{t} \longrightarrow \vec{t}_{de}=\text{diag}\{q,q,q\}\vec{t},
\end{align}
so that its affine decomposition is $A_{\mathcal{E}_q^{de}}=\text{diag}\{q,q,q\}$, $\vec{b}_{\mathcal{E}_q^{de}}=\vec{0}$. Therefore, its action over a Bell state gives the Werner state~\eqref{eq:Werner_state}, which in the Fano form is expressed as 
\begin{equation}
\mathcal{E}^{de}_{q_A} \otimes \mathcal{E}^{de}_{q_B} \left(\left|\Phi_k\right>\!\left<\Phi_k\right|\right) = \frac{1}{4}\left[\mathbb{1}_4+\sum_{i=1}^3 [\mathbb{w}_k]_{i,i}  q_A q_B \, \sigma_i \otimes \sigma_i\right].
\end{equation}

On the other hand, the Fano form corresponding to the right-hand side in Eq. \eqref{eq:calvacanti_resource} is
\begin{align}\label{eq:cavalcantiFanoForm}
    \rho_{ad}^{AB}=\frac{1}{4}\left\{\mathbb{1}_4+(1-p) \vec{k} \cdot \vec{\sigma} \otimes \mathbb{1}-(1-p) \mathbb{1} \otimes \vec{k} \cdot \vec{\sigma}+ \sum^3_{i=1} \left[p \mathbb{w}_1+(1-p) C\right]_{i,i} \, \sigma_i \otimes \sigma_i\right\}
\end{align}
being $C=\text{diag}\{0,0,-1\}$. 
This state can be seen as the result of local noises described by two amplitude-damping channels, $\mathcal{E}_q^{ad}$ and $\mathcal{E}_q^{mad}$, acting over the Bell state $\left|\Phi_1\right>\!\left<\Phi_1\right|$. Specifically, the channel $\mathcal{E}_q^{ad}$ stands for the usual form of the amplitude-damping channel, whose affine decomposition is
\begin{align}
    \vec{t} \longrightarrow \vec{t}_{ad}=\text{diag}\{\sqrt{q},\sqrt{q},q\}\vec{t} + (1-q)\vec{k},
\end{align}
being $\vec{k}=(0,0,1)^\intercal$. The map $\mathcal{E}_q^{mad}$ is the \textit{mirrored} amplitude-damping channel:
\begin{align}
    \vec{t} \longrightarrow \vec{t}_{mad}=\text{diag}\{\sqrt{q},\sqrt{q},q\}\vec{t} - (1-q)\vec{k}.
\end{align}
The only difference between these channels is that as the noise parameter $q$ goes to $0$, the images' channel concentrates around $-\vec{k}$, instead of $\vec{k}$ as in the case of the usual amplitude-damping. The action of these channels has been experimentally implemented, for instance, in Ref. \cite{Knoll2018}.
We can see that if $A_{\mathcal{E}}=A_{\mathcal{F}}=\text{diag}\{\sqrt{p},\sqrt{p},p\}$, $\vec{b}_{\mathcal{E}}=(1-p)\vec{k}$ and $\vec{b}_{\mathcal{F}}=-(1-p)\vec{k}$, Eq. \eqref{eq:noiseschannels} and Eq. \eqref{eq:cavalcantiFanoForm} coincide.

Finally, following the same reasoning as before, we can see that the Fano form of the resource state in Eq. \eqref{eq:twoADCresource} can be written as:
\begin{align}\label{eq:ADCFanoForm}
    \Tilde{\rho}_{ad}^{AB}=\frac{1}{4}\left\{\mathbb{1}_4+(1-p) \vec{k} \cdot \vec{\sigma} \otimes \mathbb{1}+(1-p) \mathbb{1} \otimes \vec{k} \cdot \vec{\sigma}+ \sum^3_{i=1} \left[p \mathbb{w}_1+(1-p)(2p-1) C\right]_{i,i} \sigma_i \otimes \sigma_i\right\}.
\end{align}

\section{Unconditionality of the optimal Bob rotations of the standard protocol with Werner state resource}
\label{app:Werneruosp}

The correlation matrix of the Werner state 
\begin{align}
    \rho^{AB}_\mathrm{W} = p\ket{\Phi_k}\bra{\Phi_k} + \frac{(1-p)}{4}\mathbb{1}
\end{align} 
can be expressed as $\mathbb{r}_\mathrm{W} = p \mw_k$. 
Consequently, the Bloch vector of the Bob state is given by $\vect_i =  p \mR_i \mw_k \mw_i \vect$. This leads to the result
\begin{align}
    \avgd = \sum^4_{i=1} \frac{1}{4} \left\langle d\left(p \mR_i \mw_k \mw_i \vect, \vect\right) \right\rangle\geq \sum^4_{i=1} \frac{1}{4} \left\langle d\left(p  \vect, \vect\right) \right\rangle,
\end{align}
where we have utilized the faithfulness property of the function $d$. Thus, the optimal universal rotations for Bob, Eq. \eqref{eq:opt_rot_werner} of the main article, can be directly derived from the condition  $\mathbb{R}_i^\mathrm{opt} \mw_k \mw_i = \mathbb{1}$. 

\section{Optimal Bob's rotations for fidelity in the standard teleportation protocol} \label{app:BobRotFid}

The corresponding output states $\rho_i^B$ of the standard teleportation protocol, described in Sec. \ref{sec:standard}, are given by 
\begin{equation}
\label{eq:outputState}
	\rho^{B}_i =U_i \rho^{B|i} U_i^\dagger=\frac{1}{2}(\mathbb{1}+\vec{t}_i \cdot \vec{\sigma}), \ \text{and } \vec{t}_i=\mR_i\vec{t}^{\,B|i}, 
\end{equation}
with
\begin{align}
    \rho^{B|i} &= \frac{1}{2} \left(\Id + \vect^{\,B|i} \cdot \vecsigma \right),  &
    \vect^{\,B|i}&=\frac{\vec{r}^B+(\mathbb{w}_i \mathbb{r})^\intercal \vec{t}}{4 p_i}, &
    p_i&=\frac{1+ \vec{t}\cdot (\mathbb{w}_i\vec{r}^A)}{4}.
\end{align} 
These states $\rho_i^B$ are dependent on Bob's rotations denoted by $\mR_i$. On the other hand, the average fidelity of teleportation (Eq. \eqref{eq:avFid}) for the standard protocol is given by \cite{Horodecki1996a}
\begin{align}
    \braket{\bar{F}} = \frac{1}{2}\left(1+\frac{1}{4\pi}\int_{S} dS \ \vec{t} \cdot \mathbb{A} \vec{t}\right)=\frac{1}{2}\left(1+ \frac{1}{3} \Tr \ \mathbb{A} \right)  \label{eq:optimalRots}
\end{align}
being $\mathbb{A}=\frac{1}{4}\sum_{i=1}^4 \mathbb{R}_i \mathbb{r}^{\intercal} \mw_i^{\intercal}$. 

Let us focus now on finding those Bob's operations maximizing Eq. \eqref{eq:optimalRots}. By using the canonical decomposition of the resource state $\rho^{AB}$ \cite{Horodecki1996a}, we have that its correlation matrix can be written as $\mathbb{r}= \mathbb{O}_1 \mathbb{r}_d \mathbb{O}_2^{\intercal}$, where $\mathbb{r}_d$ is a diagonal matrix and $\{\mathbb{O}_j\}_{j=1}^2$ are rotation matrices. Bob's rotations maximizing the average fidelity of teleportation have the form $\mathbb{R}_{i}^{\text{opt}}=-\mathbb{w}_i \mathbb{O}$ with $\mathbb{O}$ another rotation matrix, and $\mw_i$ the correlation matrix in Eqs. \eqref{eq:BellstatesCorrMat} and  \eqref{eq:BellstatesCorrMat1}, see Ref.~\cite{Horodecki1996}. Thus, the optimization problem is reduced to
\begin{align}
    \braket{\bar{F}}_{\text{max}}=\max_{\mathbb{O}}\frac{1}{2}\left(1- \frac{1}{3} \Tr \ \mathbb{r}^\intercal \mathbb{O} \right)=\max_{\mathbb{O}}\frac{1}{2}\left(1- \frac{1}{3} \Tr \  \mathbb{r}_d \mathbb{O}_1^{\intercal} \mathbb{O} \mathbb{O}_2  \right) = \max_{\mathbb{O}'}\frac{1}{2}\left(1- \frac{1}{3} \Tr \  \mathbb{r}_d \mathbb{O}'  \right), \label{eq:optimalEq}
\end{align}
which is equivalent to calculating the maximal average fidelity in the standard teleportation protocol for a Bell diagonal resource state with correlation matrix $\mathbb{r}_d=\text{diag} (r_1,r_2,r_3)$. We arrive therefore to:
\begin{align}\label{eq:avFidSQTBell}
    \braket{\bar{F}}_{\text{max}} = \max_i \left\{ \ \lambda_i (\{r_1,r_2,r_3\}) \  \right\}
\end{align}
with $\lambda_i (\{r_1,r_2,r_3\})$ the eigenvalues of the matrix $\frac{1}{4}\left(\mathbb{1}_4 + \sum_{i=1}^3 r_i \sigma_i \otimes \sigma_i \right)$. To reach the maximum in Eq. \eqref{eq:avFidSQTBell}, it is necessary to take $\mathbb{O}'_{\text{opt}}=-\mathbb{w}_l$, being $l = \text{argmax} \{\lambda_i\}_i$, in Eq. \eqref{eq:optimalEq}. The optimal Bob's rotation in Eq. \eqref{eq:optimalRots} are, therefore, of the form $\mathbb{R}_{i}^{\text{opt}}=\mathbb{w}_i \mathbb{O}_1\mathbb{w}_l\mathbb{0}_2^\intercal$. Note additionally that the optimal Bob's rotations are contingent only on the correlation matrix $\mathbb{r}$.

Note that in the case of the resource state resulting from the amplitude-damping channel and its mirrored counterpart, as given in Eq. \eqref{eq:calvacanti_resource},
\begin{align*}
    \rho^{AB}_{ad} =  p\ket{\Phi_1}\bra{\Phi_1} + (1-p)\ket{01}\bra{01}
\end{align*}
we have that $\mathbb{O}_i=\mathbb{1}$ ($i=1,2$) and the corresponding correlation matrix is $\mathbb{r}=\text{diag}(p,-p,2p-1)$. Thus, in this case, $l = \text{argmax} \{\lambda_1=\frac{1-p}{2}, \lambda_2=0, \lambda_3=\frac{1-p}{2},\lambda_4=p\}$ and we arrive at Eq. \eqref{eq:opt_Bob_rotations_calvacanti} (where the eigenvalues index is in accordance with Eqs. \eqref{eq:Bellstates} and \eqref{eq:Bellstates1}). Finally, for the last noise model, see Eq. \eqref{eq:twoADCresource}, we have that Eq. \eqref{eq:ADCFanoForm} implies that the correlation matrix in this case is $\Tilde{\mathbb{r}}=\text{diag}(p,-p,p^2+(1-p)^2)$. Following the same reasoning as before, we have that $l=\text{argmax} \{\lambda_i\}_i =1$ for all $p$.

\section{Average distance of teleportation and negativity for the considered examples} \label{app:averDistExamples}

\begin{figure*}[htb]\centering %  figure placement: here, top, bottom, or page\centering
	\includegraphics[width=\textwidth]{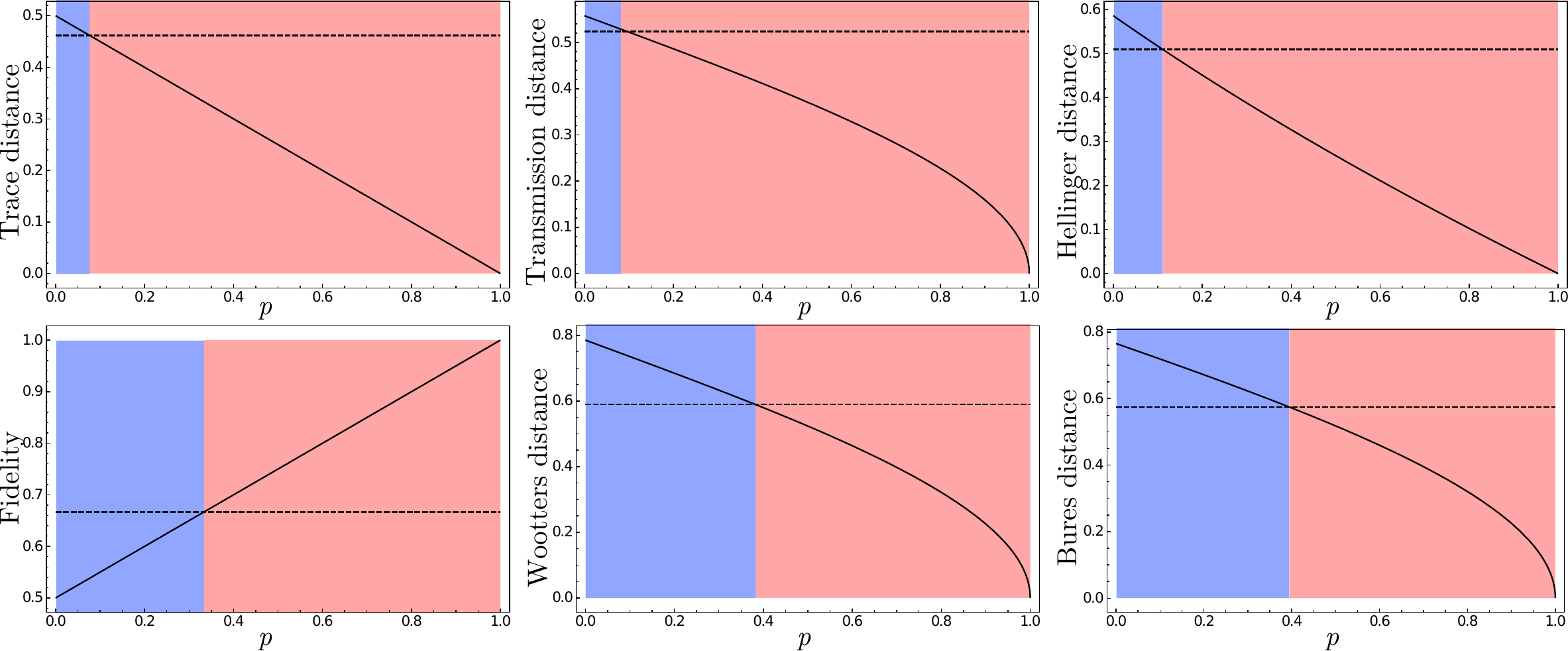}  
	\caption{How different distinguishability measures certify the protocol as either classical (blue) or quantum (red) in terms of the parameter $p$ of the Werner resource state. The solid line represents the average teleportation distance ($\avgd$), while the dashed line marks the classical threshold ($\avgd_\cl$).}
 \label{fig:avg_werner}
\end{figure*}

Fig.~\ref{fig:avg_werner} visually captures the influence of the parameter $p$ in the Werner state on protocol certification for distinguishability measures listed in Figure \ref{fig:barras}.(a) of the main text. The solid line represents the average teleportation distance ($\avgd$), while the dashed line marks the classical threshold ($\avgd_\cl$). The blue region indicates classical certification ($\avgd \geq \avgd_\cl$), and the red region signifies quantum certification ($\avgd < \avgd_\cl$). In particular, for fidelity, the inequalities are reversed.

\begin{figure*}[htb]\centering %  figure placement: here, top, bottom, or page\centering
	\includegraphics[width=\textwidth]{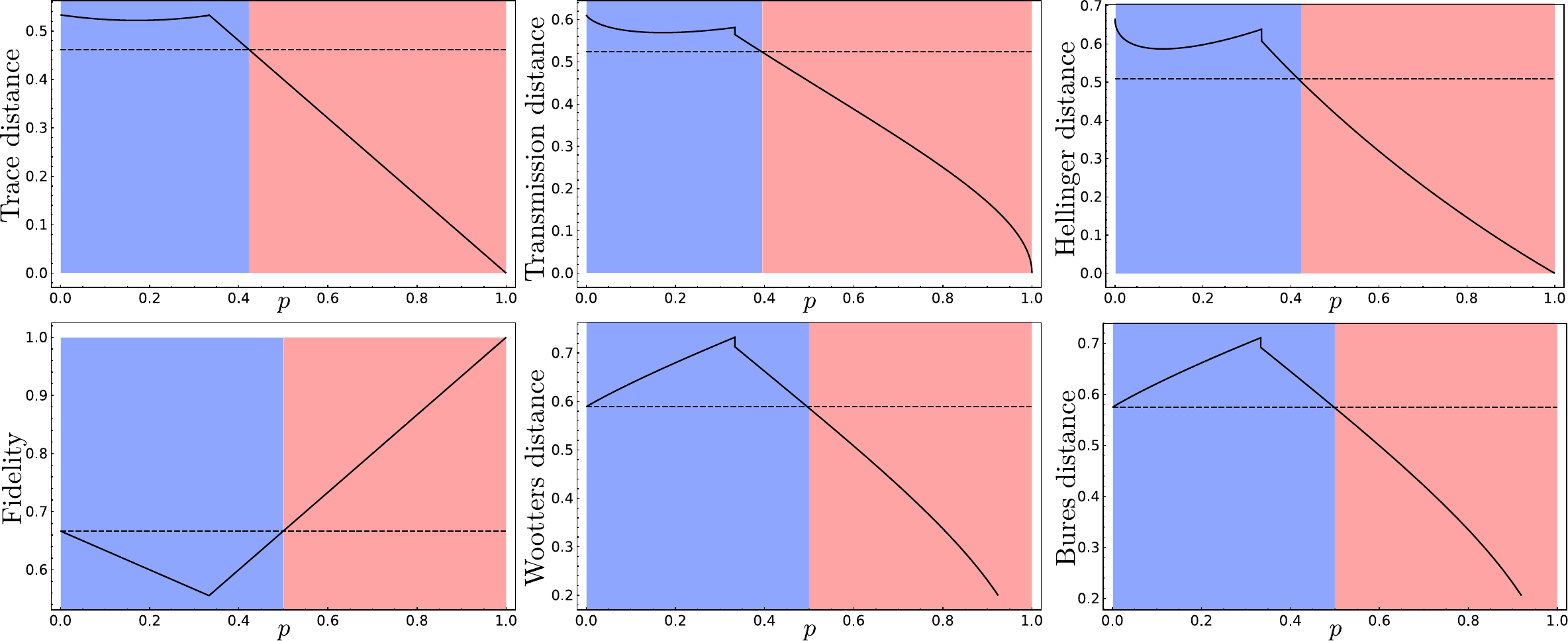}  
	\caption{How different distinguishability measures certify the protocol as either classical (blue) or quantum (red) in terms of the parameter $p$ of the resource state given in Eq. \eqref{eq:calvacanti_resource}. The solid line represents the average teleportation distance ($\avgd$), while the dashed line marks the classical threshold ($\avgd_\cl$).}
 \label{fig:avg_no_werner}
\end{figure*}

Similarly, Fig. \ref{fig:avg_no_werner} mirrors this analysis but focuses on the resource state described in Eq. \eqref{eq:calvacanti_resource}. 
We can see that all distance measures except fidelity exhibit a discontinuity point at $p=1/3$, where the optimal operations (Eq. \eqref{eq:opt_Bob_rotations_calvacanti}) change. This behavior is expected because Bob's operations are optimal for fidelity. Therefore, the discontinuity point in Fig. \ref{fig:avg_no_werner} means that each distance measure has a different set of optimal operations, allowing us to conclude that this protocol is \textit{conditionally optimal}, as defined at the beginning of Sec. \ref{sec:cavalcantiprot}. Notably, none of these considered distance measures give rise to an average teleportation distance monotonically increasing or decreasing with the noise parameter $p$.

\begin{figure*}[htb]\centering %  figure placement: here, top, bottom, or page\centering
	\includegraphics[width=\textwidth]{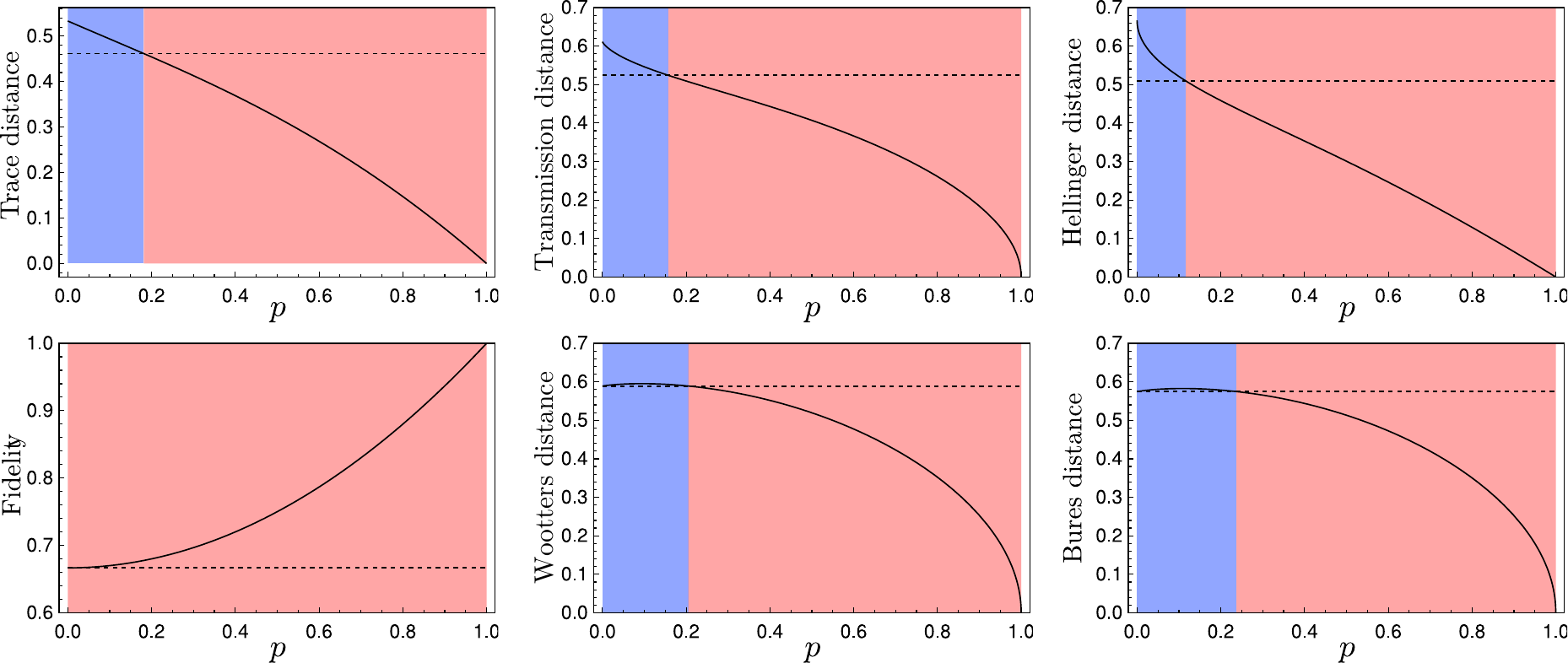}  
	\caption{Average teleportation distance (solid line, $\avgd$) for the distinguishability measures listed in Table \ref{tab:table_of_d_min}, as functions of the parameter $p$ of the resource state defined in Equation \eqref{eq:twoADCresource}. The dashed line indicates the classical limit ($\avgd_\cl$). The blue ranges correspond to classical certification while red ranges indicate quantum certification.}
 \label{fig:avg_two_adc}
\end{figure*}

The average teleportation distances corresponding to the last considered noise model in Sec. \ref{sec:twoADC}, leading to the resource state in Eq. \eqref{eq:twoADCresource}, are shown in Fig. \ref{fig:avg_two_adc}. As we can see, Wootters and Bures distances are the only distinguishability measures resulting in non-monotonic quantities with $p$.

Additionally, we obtain the resource states' negativity ($\N$) as a measure of their entanglement. In general, the  negativity is defined as follows~\cite{Vidal2002}
\begin{equation}
    \N(\rho^{AB}) = \frac{\|{\rho^{AB}}^{T_A}\|_1-1}{2},
\end{equation}
where $\|O\|_1 = \Tr|O|$ and ${\rho^{AB}}^{T_A}$ indicates partial transpose of 
$\rho^{AB}$  with respect to subsystem $A$. 
Specifically, for a Werner state, Eq. \eqref{eq:Werner_state}, is
\begin{equation}
    \label{eq:neg_werner}
    \N(\rho^{AB}_{\mathrm{W}}) =\begin{cases} 
                            0 & 0\leq p\leq \frac{1}{3} \\
                            \frac{3 p-1}{4} & \frac{1}{3} < p \leq 1. 
                        \end{cases}
\end{equation}
For the resource state given in Eq. \eqref{eq:calvacanti_resource} is
\begin{equation}
    \label{eq:neg_calvacanti}
    \N(\rho^{AB}_{ad}) =\left|\frac{1 - p - \sqrt{1 - 2 p + 2 p^2}}{2}\right|,
\end{equation}
whereas for the resource state Eq. \eqref{eq:twoADCresource} is
\begin{equation}
    \label{eq:neg_two_adc}
    \N(\tilde{\rho}^{AB}_{ad}) =\frac{p^2}{2}.
\end{equation}
Note that in Figures \ref{fig:barras} (a), (b), and (c), we plotted normalized negativity curves.

\end{document}